\documentclass[prl,final,twocolumn,showpacs,showkeys]{revtex4}
\usepackage{amsmath}
\usepackage{graphicx}
\usepackage{amsfonts}
\usepackage{amssymb}
\begin{document}

\title{Quantum logic gates for coupled superconducting phase qubits}

\author{Frederick W. Strauch}
\email[Electronic address: ]{fstrauch@physics.umd.edu}
\author{Philip R. Johnson}
\author{Alex J. Dragt}
\author{C. J. Lobb}
\author{J. R. Anderson}
\author{F. C. Wellstood}
\affiliation{Department of Physics, University of Maryland,\\College Park, Maryland 20742-4111}

\date{\today}

\begin{abstract}
Based on a quantum analysis of two capacitively coupled current-biased Josephson junctions, we propose two fundamental two-qubit quantum logic gates. Each of these gates, when supplemented by single-qubit operations, is sufficient for universal quantum computation.  Numerical solutions of the time-dependent Schr{\"o}dinger equation demonstrate that these operations can be performed with good fidelity.  
\end{abstract} 
\maketitle
The current-biased Josephson junction is an easily fabricated device with great promise as a scalable solid-state qubit \cite{Ramos2001}, as demonstrated by the recent observations of Rabi oscillations \cite{Martinis2002,Yu2002}. This phase qubit is controlled through manipulation of the bias currents and application of microwave pulses resonant with the energy level splitting \cite{Martinis2002}.  

In this Letter we analyze the quantum dynamics of two coupled phase qubits. (The classical dynamics of this system has also been studied recently \cite{Valkering2000}).  We identify two quantum logic gates that, together with single-qubit operations, provide all necessary ingredients for a universal quantum computer.  We perform full dynamical simulations of these gates through numerical integration of the time-dependent Schr{\"o}dinger equation.  These two-qubit operations may be experimentally probed with the methods already used to observe single junction Rabi oscillations \cite{Martinis2002,Yu2002}.  Such experiments are of fundamental importance: the successful demonstration of macroscopic quantum entanglement holds profound implications for the universal validity of quantum mechanics \cite{Leggett80}.  Important progress toward this goal are the temporal oscillations of coupled charge qubits \cite{Pashkin2003} and spectroscopic measurements \cite{Berkley2003s} on the system considered here.  Finally, our methods are applicable to the other promising superconducting proposals based on charge, flux, and hybrid realizations \cite{Nakamura99}.

\begin{figure}
\begin{center}
\includegraphics{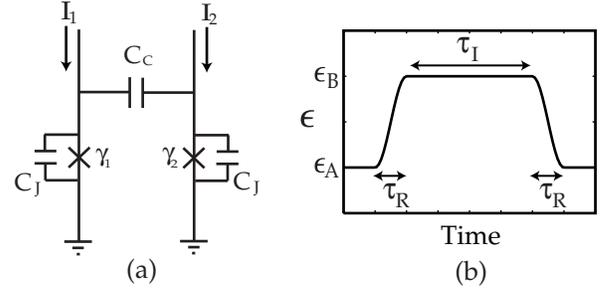}
\caption{Capacitively coupled Josephson junctions: (a) ideal circuit diagram, (possible experimental parameters include $C_J = 6$ pF, $C_C = 60.6$ fF, and $I_c = 21 \mu$A, $J_0 = 0.988$, $\omega_0/2\pi = 6.45$ GHz); (b) time-dependent ramp of bias currents, specified through the detuning $\epsilon$ (see text).}
\label{ccjj}
\end{center}
\end{figure}

Figure \ref{ccjj}(a) shows the circuit diagram of our coupled qubits.  
Each junction has characteristic capacitance $C_J$ and critical current $I_c$, and they are coupled by capacitance $C_C$.  The two degrees of freedom of this system are the phase differences $\gamma_1$ and $\gamma_2$, with dynamics governed by the Hamiltonian \cite{Johnson2002}
\begin{equation}
\begin{array}{ll}
H =& 4E_{C}(1+\zeta)^{-1}\hbar^{-2} (p_1^2 + p_2^2 + 2\zeta p_1 p_2) \\
&- E_J(\cos\gamma_1 + J_1\gamma_1 + \cos\gamma_2 + J_2\gamma_2).  
\end{array}
\label{letter1}
\end{equation}
Here we have employed the charging and Josephson energies $E_C=e^2/2C_J$ and $E_J=\hbar I_c/2e$, the normalized bias currents $J_1=I_1/I_c$, $J_2=I_2/I_c$, and the dimensionless coupling parameter $\zeta=C_C/(C_C+C_J)$.  

This coupling scheme has been recently analyzed \cite{Johnson2002,Blais2003,Ramos2003} and results in a system with easily tuned energy levels and adjustable effective coupling.  While $\zeta$ is typically fixed by fabrication, the energy levels and the effective coupling of the associated eigenstates are under experimental control through $J_1$ and $J_2$.  As shown below, the two junctions are decoupled for $J_1$ and $J_2$ sufficiently different, but if $J_1$ and $J_2$ are related in certain ways, the junctions are maximally coupled.  To illustrate this method of control, we define a reference bias current $J_0$ and consider the variation of $J_1$ and $J_2$ through a detuning parameter $\epsilon$:
\begin{equation}
\begin{array}{ll}
\sqrt{1-J_1} =& \sqrt{1-J_0}(1+\epsilon) \\
\sqrt{1-J_2} =& \sqrt{1-J_0}(1-\epsilon).
\end{array}
\label{letter2}
\end{equation}

Quantum logic gates are implemented by varying $\epsilon$ with time as shown in Fig.~\ref{ccjj}(b).  This ramps the bias currents, moving the system smoothly (with ramp time $\tau_R$) from $\epsilon_A$, where the eigenstates are essentially unentangled, to $\epsilon_B$, where the eigenstates are maximally entangled.  Entangling evolution is then allowed to occur for an interaction time $\tau_I$, after which the system is ramped back to $\epsilon_A$.  

For analysis, we use the energy scale $\hbar \omega_0 = \sqrt{8E_C E_J}(1-J_0^2)^{1/4}$ ($\omega_0/2\pi$ is the classical plasma frequency of a single junction) and the effective number $N_s$ of single junction (metastable) energy levels \cite{Johnson2002}, 
\begin{equation}
N_s = \frac{2^{3/4}}{3}\left(\frac{E_J}{E_C}\right)^{1/2}(1-J_0)^{5/4}.
\label{letter3}
\end{equation}

\begin{figure}
\begin{center}
\includegraphics{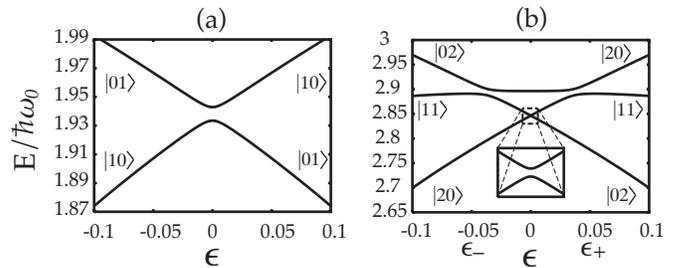}
\caption{Normalized energy levels, $N_s=4$, $\zeta = 0.01$, as a function of detuning parameter $\epsilon$: (a) the energy levels $E_1(\epsilon)$ and $E_2(\epsilon)$ with avoided level crossing at $\epsilon=0$; (b) the energy levels $E_3(\epsilon)$ through $E_5(\epsilon)$ with avoided crossings of $4$ and $5$ at $\epsilon_{\pm} \approx \pm 0.04$, and $3$ and $4$ at $\epsilon=0$.  The ground state energy $E_0$ (not shown) is approximately $0.981 \hbar \omega_0$.  The natural two-qubit states of this system are $|0;\epsilon_A)$, $|1;\epsilon_A)$, $|2;\epsilon_A)$ and $|4;\epsilon_A)$, with $\epsilon_A = -0.1$ (see text).}

    \label{energy}
  \end{center}
 \end{figure}

After choosing a fixed coupling $\zeta$, gate design requires the identification of suitable $N_s$, $\epsilon_A$, $\epsilon_B$, $\tau_R$, and $\tau_I$.  Figure \ref{energy} shows the relevant energy levels of $H$ as a function of $\epsilon$ for the physically interesting case $\zeta = 0.01$ and $N_s=4$, and with the potential energy minimum subtracted off.  The energy levels $E_n(\epsilon)$ and their associated two-junction eigenstates $|n;\epsilon)$ were computed using the method of complex scaling \cite{Johnson2002,Yaris78} applied to the cubic approximation \cite{Leggett87} of $H$.  

In general, each energy state $|n;\epsilon)$ is an entangled superposition of the product states $|jk;\epsilon\rangle = |j;\epsilon\rangle \otimes |k;-\epsilon\rangle$, where $|j; \epsilon\rangle$ are energy states of an isolated junction with normalized bias current $J_1$.  (The ``round'' and ``angular'' brackets distinguish the coupled and uncoupled bases, respectively.) However, for $|\epsilon| > 0.1$, the energy states are essentially unentangled and well approximated by the product states, which are used to label the corresponding energy levels in Fig.~\ref{energy}.  Thus for $\epsilon_A=-0.1$ we find that the eigenstates satisfy the relations $|1;\epsilon_A) \cong |10;\epsilon_A\rangle$, $|2;\epsilon_A) \cong |01;\epsilon_A\rangle$, and $|4;\epsilon_A) \cong |11;\epsilon_A\rangle$.  The ground state $|0;\epsilon) \cong |00;\epsilon\rangle$, not shown, is essentially unentangled for all $\epsilon$.  We choose these states for our two-qubit basis.  In addition, there are the auxiliary states $|3;\epsilon_A) \cong |20;\epsilon_A\rangle$ and $|5;\epsilon_A) \cong |02;\epsilon_A\rangle$.  

\begin{figure}
  \begin{center}
\includegraphics{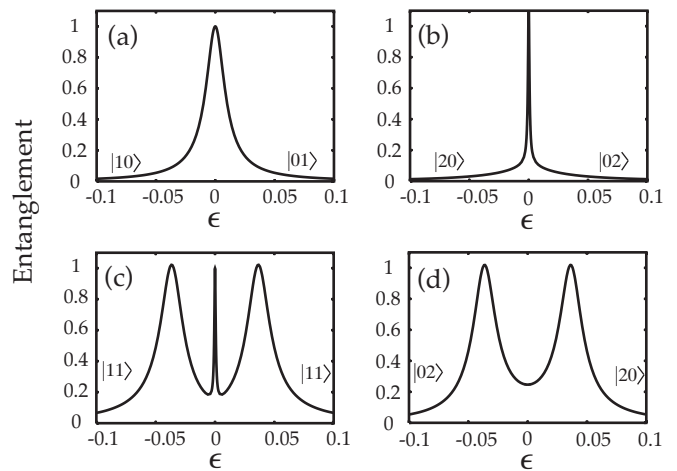}
    \caption{Entanglement, $N_s=4$, $\zeta = 0.01$:  The entanglement of the energy eigenstates as a function of detuning parameter $\epsilon$: (a) $|1;\epsilon)$; (b) $|3;\epsilon)$; (c) $|4;\epsilon)$; and (d) $|5;\epsilon)$.}
  \label{entropy}
  \end{center}
 \end{figure}

For $\epsilon$ near $\epsilon_{\pm} \cong \pm 0.04$ and $0$, where avoided level crossings occur, we find significant entanglement.  Figure \ref{entropy} shows the entanglement of the states $|n;\epsilon)$, with $n=1, 3, 4, 5$, as a function of $\epsilon$. (The entanglement of states $1$ and $2$ are nearly identical.)  The entanglement is given in ebits \cite{Bennettetal96}: a state with one ebit entanglement is a maximally entangled two-qubit state.  The gates constructed below use this entanglement to perform two-qubit operations.  

It is important to account for the nonqubit states $|02;\epsilon\rangle$ and $|20;\epsilon\rangle$---poorly designed interactions will result in unwanted evolution of $|11;\epsilon\rangle$ into these auxiliary levels, an effect called leakage.   However, we show below that at both $\epsilon_B=0$ and $\epsilon_B=\epsilon_-$ the state $|11;\epsilon_B\rangle$ is a superposition of only two of the energy eigenstates.  Therefore, its time evolution is of the form $|\langle 11;\epsilon_B|e^{-iH\tau_I/\hbar}|11;\epsilon_B\rangle|^2 = a+b\cos^2(\Omega \tau_I)$, with $a+b=1$.  Choosing $\tau_I = k \pi/\Omega$, where $k$ is an integer, ensures that the oscillation of $|11;\epsilon\rangle$ with the auxiliary states completes $k$ full cycles, minimizing leakage.  This procedure is similar to operations performed in ion traps \cite{Cirac95}. 

If we allow $|11;\epsilon\rangle$ to evolve through states $|02;\epsilon\rangle$ and $|20;\epsilon\rangle$, we must consider another possibility for error: the tunneling rates of these auxiliary states are higher than that of $|11;\epsilon\rangle$.  As these tunneling rates are all proportial to $e^{-36 N_s /5}$ \cite{Leggett87}, we minimize this error by choosing a large $N_s$.  For example, to achieve a fidelity greater than $0.999$, we find that $N_s \ge 4$ is necessary.  We note that proper operation requires the ramp time $\tau_R$ to be adiabatic with respect to single junction energy level spacings, yet nonadiabatic with respect to the coupled energy level splittings.  This leads to our choice $\zeta=0.01$.


Before analyzing gate dynamics, we look more closely at the eigenstates at $\epsilon=0$ and $\epsilon_{\pm}$.  These can be understood perturbatively and identified with degeneracies in the combined spectrum of two uncoupled Josephson junctions.  For example, with $\zeta=0$ at $\epsilon=0$ the states $|01;0\rangle$ and $|10;0\rangle$ are degenerate.  Coupling splits this degeneracy \cite{Blais2003}, and yields the true eigenstates
\begin{equation}
\begin{array}{lcl}
|1;0) = 2^{-1/2}(|01;0\rangle - |10;0\rangle)\\
|2;0) = 2^{-1/2}(|01;0\rangle + |10;0\rangle).
\end{array}
\label{letter5}
\end{equation}
These states (Fig.~\ref{entropy}(a)) are maximally entangled qubits.  Similarly, states $|02;0\rangle$ and $|20;0\rangle$ are degenerate with $\zeta=0$. With nonzero coupling, however, the energy splitting in this case is much smaller [see the inset to Fig~\ref{energy}(b)].  Here, a three-state analysis is required, with the nearly degenerate $|11;0\rangle$ mediating the coupling.  The eigenstates are approximately
\begin{equation}
\begin{array}{lcl}
|3;0) = 2^{-1/2}\cos\theta(|02;0\rangle + |20;0\rangle) - \sin\theta |11;0\rangle \\
|4;0) = 2^{-1/2}(|02;0\rangle - |20;0\rangle)\\
|5;0) = 2^{-1/2}\sin\theta(|02;0\rangle + |20;0\rangle) + \cos\theta |11;0\rangle,
\end{array}
\label{letter6}
\end{equation}
with $\theta \cong 0.185$.  In accord with Fig.~\ref{entropy}(b-d), each of these states is substantially entangled at $\epsilon=0$ \cite{Bennettetal96}, while $|11;0\rangle$ is a superposition of $|3;0)$ and $|5;0)$.

Finally, perturbation theory shows that the off-symmetry degeneracies occur for $\epsilon_{\pm} \cong \pm 5/36N_s$.  For example, at $\epsilon=\epsilon_{-}$ with $\zeta=0$ states $|02;\epsilon_-\rangle$ and $|11;\epsilon_-\rangle$ are degenerate.  Coupling leads to the true eigenstates
\begin{equation}
\begin{array}{lcl}
|4;\epsilon_-) = 2^{-1/2}(|02;\epsilon_-\rangle - |11;\epsilon_-\rangle) \\
|5;\epsilon_-) = 2^{-1/2}(|02;\epsilon_-\rangle + |11;\epsilon_-\rangle). 
\end{array}
\label{letter8}
\end{equation}
As seen in Fig.~\ref{entropy}(c,d), both eigenstates have an entanglement of 1 ebit.  Further, we see that $|11;\epsilon_-\rangle$ is a superposition of $|4;\epsilon_-)$ and $|5;\epsilon_-)$. 

We now describe the two-qubit gate that uses the entanglement at $\epsilon_-$.  
As noted above, if the state $|11;\epsilon_-\rangle$ is prepared, its time evolution (for fixed $\epsilon$) will be oscillatory.  Letting $\tau_I = 2\pi\hbar/[E_{5}(\epsilon_-) - E_{4}(\epsilon_-)] \cong \sqrt{2}\pi/\zeta \omega_0$, $|11;\epsilon_-\rangle$ performs a complete oscillation, while picking up an overall controlled phase.  The remaining states also evolve dynamical phases, which can be factored out as one-qubit gates by letting $U_1 = e^{i\alpha_1}R_z(\alpha_2)\otimes~R_z(\alpha_3)e^{-iH\tau_I/\hbar}$, with $R_z(\theta) = e^{-i\theta \sigma_z/2}$ ($\sigma_z$ is a Pauli matrix).  Here $\alpha_1 = [E_1(\epsilon_-)+E_2(\epsilon_-)]\tau_I/2\hbar$, $\alpha_2 = [E_1(\epsilon_-)-E_0(\epsilon_-)]\tau_I/\hbar$, and $\alpha_3 = [E_2(\epsilon_-)-E_0(\epsilon_-)]\tau_I/\hbar$.  In our two-qubit basis, this operation is the controlled-phase gate
\begin{equation}
U_1 = 
\left(
\begin{array}{cccc}
1 & 0 & 0 & 0 \\
0 & 1 & 0 & 0 \\
0 & 0 & 1 & 0 \\
0 & 0 & 0 & e^{-i\phi}
\end{array}
\right)
\label{letter10}
\end{equation}
with $\phi = [E_4(\epsilon_-)+E_0(\epsilon_-)-E_1(\epsilon_-)-E_2(\epsilon_-)]\tau_I/\hbar$.   For $N_s=4$ and $\zeta=0.01$, we find $\phi \cong 1.02\pi$, thus this gate is approximately the controlled-$Z$ gate \cite{Nielsen2000}.

We can also use the entanglement at $\epsilon_B=0$ for quantum logic.  From the dynamics of $|11;0\rangle$ [implied by Eq.~(\ref{letter6})], we let the interaction time be $\tau_I=2\pi k \hbar/[E_{5}(0)-E_{3}(0)]$, where $k$ is an integer.  From Eq.~(\ref{letter5}), however, the states $|01;0\rangle$ and $|10;0\rangle$ will also oscillate \cite{Blais2003}.  Removing one-qubit dynamical phases as above, we define $U_2 = e^{i\alpha_1}R_z(\alpha_2)\otimes~R_z(\alpha_3)e^{-iH\tau_I/\hbar}$, with $\alpha_1=[E_1(0)+E_2(0)]\tau_I/2\hbar$ and $\alpha_2=\alpha_3 = [E_1(0)+E_2(0)-2E_0(0)]\tau_I/2\hbar$.  For $U_2$ we find the swaplike gate
\begin{equation}
U_2 = \left(
\begin{array}{cccc}
1 & 0 & 0 & 0 \\
0 & \cos\theta_1 & -i\sin\theta_1 & 0 \\
0 & -i\sin\theta_1 & \cos\theta_1 & 0 \\
0 & 0 & 0 & e^{-i\theta_2}
\end{array}
\right)
\label{letter11}
\end{equation}
with swap angle $\theta_1 = [E_2(0)-E_1(0)]\tau_I/2\hbar$ and controlled-phase $\theta_2 = [E_5(0)+E_0(0)-E_1(0)-E_2(0)]\tau_I/\hbar$.  As $\theta_1$ and $\theta_2$ are in general irrational multiples of $\pi$, this gate is universal for quantum computation \cite{Lloyd95}.  For example, by tuning $J_0$ such that $N_s \cong 5.16$ and letting $k=2$, the full swap dynamics is generated, with $\theta_1=\pi/2$, $\theta_2 \cong \pi/4$, and $\tau_I \cong \pi/\zeta\omega_0$.  

Proceeding beyond this heuristic analysis, we have numerically solved the time-dependent Schr{\"o}dinger equation using split-operator unitary integration \cite{Takahashi93}.  Tunneling is incorporated through an absorbing boundary condition \cite{Kosloff86}.  Taking $\epsilon_A=-0.1$ as our initial detuning, we evolved states having initial conditions $|0;\epsilon_A)$, $|1;\epsilon_A)$, $|2;\epsilon_A)$, and $|4;\epsilon_A)$ using the the ramp function of Fig.~\ref{ccjj}(b).  While the results quoted below are in the cubic approximation, results obtained using the full Hamiltonian are only marginally different.

\begin{figure}[b]
  \begin{center}
\includegraphics{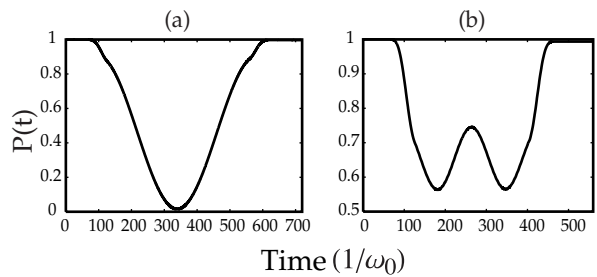}
    \caption{Dynamical evolution of state with initial condition $|4;\epsilon_A)$. The probability $P(t)=|(4;\epsilon_A|\psi(t)\rangle|^2$ is shown for (a) the phase gate $U_1$ and (b) the swaplike gate $U_2$.}
    \label{letterx1}
  \end{center}
 \end{figure}


\begin{figure}
  \begin{center}
\includegraphics{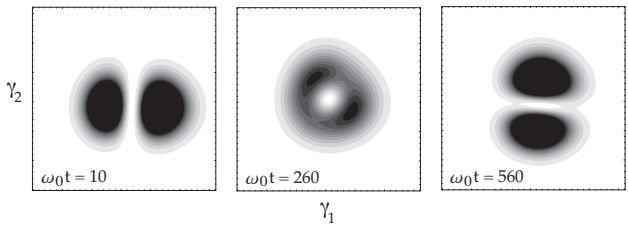}
    \caption{Dynamical evolution of state with initial condition $|1;\epsilon_A)$ under the swaplike gate.  The contours represent the numerically computed wave function (modulus squared) evolving from what is nearly $|10;\epsilon_A\rangle$ to $|01;\epsilon_A\rangle$. }
    \label{swapwfxns}
  \end{center}
 \end{figure}

The controlled-phase gate $U_1$ is simulated with $N_s=4$ and $\zeta=0.01$. Using the minimum splitting between $E_4(\epsilon)$ and $E_5(\epsilon)$ we take $\epsilon_B=-0.036$, with $\tau_R = 20\pi/\omega_0$ and $\tau_I = 434/\omega_0$.  The dynamical behavior of this gate is illustrated in Fig.~\ref{letterx1}(a), where the probability $P(t)=|(4;\epsilon_A|\psi(t)\rangle|^2$ with $|\psi(0)\rangle = |4;\epsilon_A)$ is shown.  Including all two-qubit states, we find an average gate fidelity \cite{Nielsen2002} $F=0.996$, with a leakage \cite{Fazio99} $L=0.003$.  

For the swaplike gate $U_2$, we use $N_s = 5.16$, $\zeta=0.01$, $\tau_R=20\pi/\omega_0$ and $\tau_I=278/\omega_0$.  The probability $P(t)$ is shown in Fig.~\ref{letterx1}(b), completing two oscillations ($k=2$) but with diminished amplitude due to the shift of the potential (since $|\langle 11;\epsilon_A|11;\epsilon_B \rangle|^2 < 1$).  The swap of $|10;\epsilon_A\rangle$ and $|01;\epsilon_A\rangle$ is shown in Fig.~\ref{swapwfxns}.  This gate is not quite as good as $U_1$, with a fidelity $F=0.972$ and leakage $L=0.006$.  We expect that a complete optimization of $U_1$ and $U_2$ will yield sufficient fidelity for fault-tolerant quantum computation \cite{Preskill97}. 

An experimental demonstration of these gates requires the following considerations.  First, for typical experimental parameters [see Fig.~\ref{ccjj}(a)] the plasma frequency $\omega_0/2\pi$ is near $6$ GHz.  With this value, the ramp time used above is $\tau_R \cong 1.67$ ns.  The total gate times ($\tau_I + 2\tau_R$) are $\tau_1 \cong 14.85$ ns for the phase gate and $\tau_2 \cong 10.7$ ns for the swap gate.  Pulse shapes similar to Fig.~\ref{ccjj}(b) with these times can be engineered with convential electronics.  Second, coherent operation at these time scales requires low dissipation due to the control circuit, which is possible with impedance transformers \cite{Martinis2002}.  Finally, the essential parameters controlling the gate dynamics are the energy level spacings.  Since these can be determined spectroscopically \cite{Johnson2002,Berkley2003s}, all aspects of this design are experimentally accessible.

Other important issues are noise in the bias currents and nonidentical junction parameters.  Current noise will cause fluctuations in $\epsilon$, while nonidentical junctions will have reduced symmetry about $\epsilon=0$.  The $U_2$ gate is particularly sensitive to these, as it uses both the symmetry and delicate structure of the eigenstates at $\epsilon=0$.  In contrast, since the phase gate $U_1$ operates at an off-symmetry position, its operation will be less sensitive to these sources of decoherence.

The controlled coupling of qubits can be refined by introducing a middle junction to generate entanglement between adjacent qubits \cite{Blais2003}.  Then, on a state of the form $|\Psi\rangle = c_{00}|000\rangle + c_{01}|001\rangle + c_{10}|100\rangle + c_{11}|101\rangle$, the operation $U=(U_2\otimes I)(I\otimes U_1)(U_2\otimes I)$ (where $I$ is the one-qubit identity operator) is equivalent (with single-qubit operations on the outer junctions) to performing a phase gate on the outermost qubits, leaving the central qubit in its ground state.  In this scheme the swaplike gate $U_2$ does not act on $|11;\epsilon\rangle$, so decoherence effects will be less severe than indicated above.  Furthermore, a larger coupling may be used, leading to smaller gate times.

In conclusion, we have shown how to implement two quantum logic gates in this coupled junction system.  For $U_1$, evolution through auxiliary levels outside of the two-qubit basis generates a controlled phase on state $|11;\epsilon\rangle$, while for $U_2$ an additional swap operation is performed between states $|01;\epsilon\rangle$ and $|10;\epsilon\rangle$.   Finally, we expect that a 3-junction design, using the unitary operations identified here, is an even better candidate for universal quantum computation with capacitively coupled Josephson junctions.


\begin{acknowledgments}
We would like to thank A.J. Berkley, J.M. Martinis, R. Ramos, H. Xu, and M. Gubrud for useful discussions.  This work was supported in part by the U.S. Department of Defense and the State of Maryland through the Center for Superconductivity Research.

\end{acknowledgments}

\bibliography{scqc}

\end{document}